\documentclass[a4paper,
               keeplastbox,   
               ]{jacow}
\usepackage{pdfpages,multirow,ragged2e} %
\makeatletter%
	\ifboolexpr{bool{xetex}}
	 {\renewcommand{\Gin@extensions}{.pdf,%
	                    .png,.jpg,.bmp,.pict,.tif,.psd,.mac,.sga,.tga,.gif,%
	                    .eps,.ps,%
	                    }}{}
\makeatother

\ifboolexpr{bool{xetex} or bool{luatex}} 
 {}                                      
 {\usepackage[utf8]{inputenc}}           

\usepackage[USenglish]{babel}

\ifboolexpr{bool{jacowbiblatex}}%
 {%
  \addbibresource{jacow-test.bib}
  \addbibresource{biblatex-examples.bib}
 }{}
\begin{document}

\title{Evaluation of the superconducting characteristics of multi-layer thin-film structures of ${\rm{\textbf{NbN}}}$ and ${\rm{\textbf{SiO}}}$$_2$ on pure ${\rm{\textbf{Nb}}}$ substrate}

\author{R. Katayama\thanks{ryo.katayama@kek.jp}, 
	H. Hayano, T. Kubo, T. Saeki (KEK, Ibaraki)\\
	Hayato Ito (Sokendai, Ibaraki)\\
	Y. Iwashita, H. Tongu (Kyoto ICR, Uji, Kyoto),\\
	C. Z. Antoine (CEA/IRFU, Gif-sur-Yvette),\\
	R. Ito, T. Nagata (ULVAC, Inc, Chiba)	}

\maketitle

\begin{abstract}
In recent years, it has been pointed out that the maximum accelerating gradient of a superconducting RF cavity can be increased by coating the inner surface of the cavity with a multilayer thin-film structure consisting of alternating insulating and superconducting layers. In this structure, the principal parameter that limits the performance of the cavity is the critical magnetic field or effective $H_{C1}$ at which vortices begin penetrating into the superconductor layer. This is predicted to depend on the combination of the film thickness. We made samples that have a NbN/SiO$_2$ thin-film structure on a pure Nb substrate with several layers of NbN film deposited using DC magnetron sputtering method. Here, we report the measurement results of effective $H_{C1}$ of NbN/SiO$_2$(30 nm)/Nb multilayer samples with thicknesses of NbN layers in the range from 50 nm to 800 nm by using the third-harmonic voltage method. Experimental results show that an optimum thickness exists, which increases the effective $H_{C1}$ by 23.8 \%.
\end{abstract}

\section{INTRODUCTION}
\noindent
Recently, it has been pointed out that the effective $H_{C1}$ of a superconducting RF cavity can be increased by coating the inner surface of the cavity with a multilayer thin-film structure consisting of alternate insulating and superconducting layers \cite{ref0,ref1,ref2}. Generally, the effective $H_{C1}$ of a superconducting material can be evaluated by applying an AC magnetic field to the material with a small coil and detecting the third-harmonic signal of the coil voltage. This third-harmonic signal occurs when the phase transition from the full Meissner state to the vortex-penetrating state happens. Hereafter, this method is called the third harmonic voltage method. The third harmonic measurement system has already been constructed at Kyoto University to evaluate the superconducting characteristics of samples of S-I-S structure in the low temperature region up to the temperature of liquid helium. We have verified that the effective $H_{C1}$ is enhanced for a sample with S-I-S structure that consists of NbN (200 nm) and SiO$_2$ (30 nm) formed on a pure Nb substrate that has an RRR of $>$250 \cite{ref3}\cite{ref3-2}. Our measurement results of the third harmonic voltage using AC magnetic fields of 5 kHz and an amplitude of less than 44 mT are reported in the LINAC18 proceedings. In this article, we will present new measurement results of effective $H_{C1}$ of NbN/SiO$_2$(30 nm)/Nb samples with thicknesses of the NbN layer in the range 50 nm - 800 nm using an AC magnetic field of 5 kHz with amplitude up to about 90 mT.

\section{THIRD HARMONIC MEASUREMENT}
\noindent
For the third harmonic voltage method, an AC magnetic field at a frequency of 5 kHz is generated by a coil close to the superconducting sample. The third harmonic voltage $v_3(t)=V_3\sin{\left(3\omega{t}\right)}$ induced in the coil is simultaneously measured \cite{ref4}; $\omega$ is the frequency of a sinusoidal drive current, $I_0\sin{\left(\omega{t}\right)}$ represents the current flowing in the coil, and $V_3$ is the amplitude of $v_3(t)$. If the temperature of a sample in the superconducting state is increased while the amplitude of the AC magnetic field $H_0$ is fixed, $V_3$ suddenly rises when $H_0$ exceeds the effective $H_{C1}$ of the sample at a particular temperature. In the measurement performed at Kyoto University, $H_0$ is controlled by drive current $I_0$, and the temperature dependence of the effective $H_{C1}$ is estimated from the temperature at the point when $V_3/I_0$ rises. Refer to \cite{ref3} for details of the measurement setup and flow.

\begin{figure}[!htb]
  \centering
  \includegraphics*[width=0.8\columnwidth]{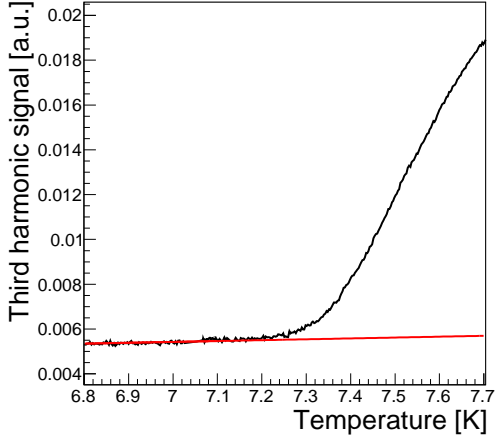}
  \caption{\label{fig1} Example of temperature dependence of third harmonic signal. The horizontal and vertical axes denote the temperature of the measured sample and the third harmonic signal used in this study, respectively.}
\end{figure}

For example, the temperature dependence of a typical third harmonic signal is shown in Fig.\ref{fig1}. Horizontal and vertical axes represent the temperature and the third harmonic signal, respectively. The red curve is the result of fitting a single polynomial function to the baseline value before the signal begins rising. The third harmonic signal is fluctuated around the baseline with a standard deviation $\sigma$, so that we determined the temperature at which the third harmonic signal suddenly rise by using a one-tailed test of a significance level of 0.013 \%.

In this study, the coil magnetic field is calibrated with the third harmonic measurement result of pure bulk Nb, assuming that the following function $F(T)$ represents the temperature dependence of the effective $H_{C1}$ of pure bulk Nb. Refer to \cite{ref3} for details of the calibration.

\begin{equation}
F(T) = \left\{ \begin{array}{ll}
  0.18\times{\left(1-(T/9.2)^2\right)} & (T < 9.2 \mathrm{K}) \\
  0 & (T > 9.2 \mathrm{K})
 \end{array} \right.
 \label{eq1}
\end{equation}

\section{DETAILS OF MEASUREMENT RESULTS}
\noindent
We have tested nine multilayer samples that consist of NbN and SiO$_2$ coated on pure bulk Nb. The pure bulk Nb substrate of the sample is pretreated with the standard electropolishing recipe for bulk Nb cavity. These multilayer samples are prepared using DC magnetron sputtering technique (by ULVAC, Inc.). The samples have a thin-film structure of NbN film of various thicknesses (50 nm, 100 nm, 150 nm, 200 nm, 250 nm, 270 nm, 300 nm, 400 nm, 800 nm) and 30-nm-thick SiO$_2$ film; the details of thin-film preparation are given elsewhere \cite{ref5}\cite{ref6}. In the third harmonic measurements, the temperature ramping rates were kept at several 0.01 K/min. The systematic error in the measured temperature was estimated to be 0.02 K owing to a thermal non-uniformity. In this study, only for 50 nm film thickness of NbN, we used the measurement data reported in the LINAC18 proceedings. 

\begin{figure}[!htb]
  \centering
  \includegraphics*[width=0.95\columnwidth]{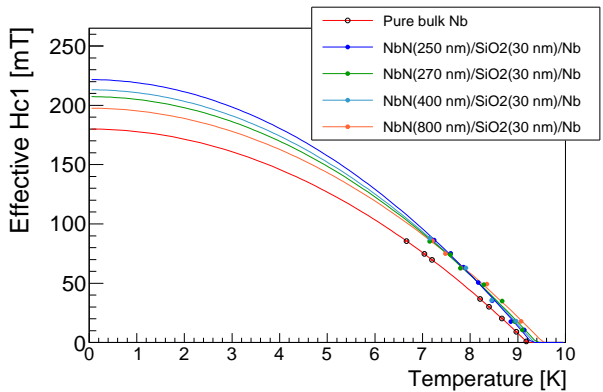}
  \caption{\label{fig2}Comparison of effective $H_{C1}$ between pure bulk Nb sample and measured NbN/SiO$_2$(30 nm)/Nb multilayer samples. The red curve represents the equation (\ref{eq1}), which is used for calibration. Other colored lines are obtained by fitting data points of NbN/SiO$_2$/Nb multilayer samples with equation (\ref{eq2}).}
\end{figure}

The analysis results of the temperature dependence of the effective $H_{C1}$ of NbN/SiO$_2$/Nb multilayer samples are depicted in Fig.\ref{fig2}. In this plot, $H_{C1}$ of pure bulk Nb used for the calibration is also plotted for comparison. The horizontal and vertical axes represent temperature and effective $H_{C1}$, respectively. An uncertainty owing to thermal non-uniformity and an error of the Nb calibration curve are assigned to each data point. The measured values of effective $H_{C1}$ of pure bulk Nb sample and the effective $H_{C1}$ of NbN/SiO$_2$(30 nm)/Nb multilayer samples are represented by the open circles and closed circles, respectively. The red curve is the theoretical curve obtained from equation (\ref{eq1}), which is used for calibration. Other colored curves are obtained by fitting the data points of the samples of NbN/SiO$_2$/Nb to the function (\ref{eq2}). It is noted that at temperatures below around 9.2 K, an S-I-S structure is formed because both the pure bulk Nb and the NbN film are in a superconducting state. Fig.\ref{fig2} thus represents the comparison between effective $H_{C1}$ of pure bulk Nb and NbN/SiO$_2$/Nb multilayer samples when the S-I-S structure exists. Consequently, we confirmed that the effective $H_{C1}$ of all NbN/SiO$_2$/Nb multilayer samples, except when the NbN thickness is 50 nm, are increased from that of bulk Nb sample. The effective $H_{C1}$ of NbN/SiO$_2$/Nb multilayer samples varies depending on the NbN film thickness.

\begin{figure}[!htb]
  \centering
  \includegraphics*[width=0.8\columnwidth]{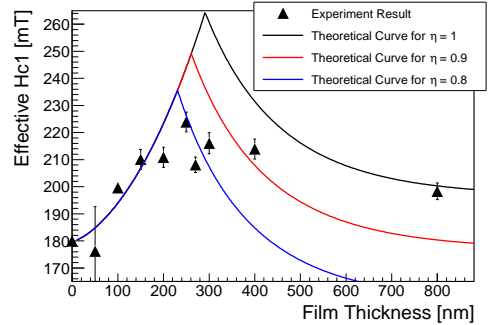}
  \caption{\label{fig3} Dependence of the effective Hc1 of NbN/SiO$_2$/Nb multilayer samples on the thickness of the NbN layer. The theoretical calculation is superimposed for comparison. Black triangles represent the measurement values of the effective $H_{C1}$ for each NbN/SiO$_2$/Nb multilayer sample.
	}
\end{figure}

The analysis result of the dependence of effective $H_{C1}$ of NbN/SiO$_2$(30 nm)/Nb multilayer samples at 0 K on the thickness of the NbN layer is shown in Fig.\ref{fig3}. The horizontal and vertical axes represent the film thickness of the NbN layer and the effective $H_{C1}$ at 0 K, respectively. The error on the vertical axis is due to a fitting error of equation (\ref{eq2}). In Fig.\ref{fig3}, theoretical calculations based on \cite{ref7} as well as experimental results are superimposed. $\eta$ (0 $< \eta <$ 1) is a parameter to indicates how the effective $H_{C1}$ of the NbN layer deteriorates owing to the effect of imperfect surfaces such as defects and surface roughness. The case where $\eta$ = 1 corresponds to an ideal smooth surface of the NbN layer; the smaller $\eta$ is, the lower the effective $H_{C1}$ of the NbN layer becomes \cite{ref8}\cite{ref9}. Experimental results clearly show that there is an optimum thickness of the NbN layer to maximize effective $H_{C1}$, which is the same as predicted. The maximum effective $H_{C1}$ value for our experiment was increased by 23.8 \% compared to pure bulk Nb. In particular, data points for up to 270 nm thickness are in good agreement with the theoretical prediction for the case of $\eta$ = 0.8. Meanwhile, data points for greater than 300 nm thicknesses of the NbN layer tend to shift toward the theoretical predictions for the case of $\eta>$ 0.8. This phenomenon is considered to be caused by the improvement in the quality of the NbN layer as the film thickness increases.

\subsection{SUMMARY}
\noindent
By using the third harmonic system constructed at Kyoto University, we evaluated the temperature dependence of effective $H_{C1}$ of samples having an S-I-S structure consisting of an NbN superconducting layer and an SiO$_2$ insulating layer (30 nm) formed on pure bulk Nb, where the thicknesses of the NbN film are in the range from 50 nm - 800 nm. Experimental results clearly showed that the effective $H_{C1}$ of all samples except for NbN film thickness of 50 nm are increased compared to that of pure bulk Nb. It is proven that an optimum film thickness exists to improve the effective $H_{C1}$ of NbN/SiO$_2$/Nb multilayer samples; this results in an increase of 23.8\% compared to that of bulk Nb. Theoretical calculations were compared to experimental results. We confirmed that experimental results are qualitatively consistent with the theoretical prediction. In particular, measured data in the range of thicknesses of the NbN layer from 50 to 270 nm showed good agreement with the theoretical calculation for $\eta=0.8$. The effective $H_{C1}$ evaluated from measured data for thickness of the NbN layer greater than 270 nm shifts to that of the theoretical prediction for $\eta >$ 0.8, owing to an improvement of the quality of NbN film. 
These results imply the possibility that the acceleration gradient of superconducting RF cavities can be effectively enhanced by controlling the multilayer thin film structure, and show the possibility of getting high-performance superconducting RF cavities with thin-film technology in mass-production consistently.

\section{ACKNOWLEDGMENTS}
\noindent
This work is supported by JSPS KAKENHI Grant Number JP19H04395, JSPS KAKENHI Grant Number JP17H04839, JSPS KAKENHI Grant Number JP26600142, Photon and Quantum Basic Research Coordinated Development Program of MEXT, Japan, Center of Innovation (COI) Program, Japan-US Research Collaboration Program, and the Collaborative Research Program of Institute for Chemical Research, Kyoto University (2019-3). 

\section{APPENDIX}
In this paper, the following function is used in Fig.\ref{fig2}:
\begin{eqnarray}
f(T) &=&  f(0)\times{\left(1-(T/T_C^\prime)^2\right)} \label{eq2}
\end{eqnarray}

%
%
\ifboolexpr{bool{jacowbiblatex}}%
	{\printbibliography}%

\begin{thebibliography}{99} 
\bibitem{ref0} A. Gurevich, ``Enhancement of rf breakdown field of superconductors by multilayer coating'', Appl. Phys. Lett. 88, 012511 (2006).
\bibitem{ref1} T. Kubo, {\it et al.}, ``Radio-frequency electromagnetic field and vortex penetration in multilayered superconductors'', Appl. Phys. Lett. 104, 032603 (2014).
\bibitem{ref2} T. Kubo, ``Multilayer coating for higher accelerating fields in superconducting radio-frequency cavities: a review of theoretical aspects'', Supercond. Sci. Tech-nol. 30, 023001 (2017).
\bibitem{ref3} R. Katayama, {\it et al.}, ``Evaluation of superconducting characteristics on the thin-film structure by NbN and Insulator coatings on pure Nb substrate'', IPAC2018 Proceedings, Vancouver, Canada.
\bibitem{ref3-2} R. Katayama, {\it et al.}, ''Precise Evaluation of Characteristic of the Multi-layer Thin-film Superconductor Consisting of NbN and Insulator on Pure Nb Substrate``, LINAC2018 Proceedings, Beijing, China.
\bibitem{ref4} Y. Mawatari, {\it et al.}, ``Critical current density and third-harmonic voltage in superconducting films'', Appl. Phys. Lett. 81, 2424 (2002).
\bibitem{ref5} R. Ito, T. Nagata, {\it et al.}, ``Development of Coating Technique for Superconducting Multilayered Structure'', IPAC2018 Proceedings, Vancouver, Canada.
\bibitem{ref6} R. Ito, T. Nagata, {\it et al.}, ``Construction of Thin-film Coating System Toward the Realization of Superconducting Multilayered Structure'', LINAC18 Proceedings, Beijing, China, TUPO050.
\bibitem{ref7} T. Kubo {\it et al.}, ''Radio-frequency electromagnetic field and vortex penetration in multi-layered super-conductors``, Appl. Phys. Lett. 104, 032603 (2014).
\bibitem{ref8} T. Kubo, ``Field limit and nano-scale surface topography of superconducting radio-frequency cavity made of extreme type II superconductor'', Progress of Theoretical and Experimental Physics, 2015, 063G01 (2015).
\bibitem{ref9} A. Gurevich and T. Kubo, ``Surface impedance and optimum surface resistance of a superconductor with an imperfect surface'', Phys. Rev. B 96, 184515 (2017).
\end{thebibliography}
	{%
	

} 
%
%


\end{document}